\documentclass[showpacs,preprintnumbers,amsmath,amssymb]{revtex4}

\usepackage{graphicx}
\usepackage{dcolumn}
\usepackage{bm}

\begin{document}
\title{Spin Alignment of Vector Mesons in Non-central $A+A$ Collisions}
\author{Zuo-Tang Liang$^1$ and Xin-Nian Wang$^{2,1}$}
\affiliation{
$^1$Department of Physics, Shandong University, Jinan, Shandong 250100,
China\\
$^2$Nuclear Science Division, MS 70R0319,
Lawrence Berkeley National Laboratory, Berkeley, California 94720}

\date{\today}

\preprint{LBNL-56659}

\begin{abstract}
We discuss the consequence of global polarization
of  the produced  quarks in non-central heavy-ion
collisions on the spin alignment of vector mesons. We show that
the alignment is quite different for different hadronization scenarios.
These results can be tested directly by measuring the
vector mesons' alignment through angular distributions of the
decay products with respect to the reaction plane.
Such angular distributions will give rise to azimuthal
anisotropy $v_2$ of the decay products in the collision frame.
Constraints provided by the data on the azimuthal anisotropy
of hadron spectra at RHIC points to a quark recombination scenario
of hadronization.
\end{abstract}


\maketitle

\vspace{10pt}


Due to the presence of a large orbital angular momentum of the
parton system produced at the early stage of non-central heavy-ion
collisions, quarks and anti-quarks are shown recently
\cite{Liang:2004ph} to be polarized in the direction opposite to
the reaction plan which is determined by the vector of
impact-parameter and the beam momentum. Such global quark
polarization should have many observable consequences such as
left-right (with respect to the quarks' spin polarization)
asymmetry in hadron spectra at large rapidity, global polarization
of thermal photons, dileptons and final hadrons with spin. The
left-right asymmetry of hadrons from the hadronization of such
globally polarized quarks effectively gives rise to a ``directed
flow''. However, it is difficult to disentangle it from the true
directed flow due to interaction between produced matter and the
spectator nucleons. Global hyperon polarization from the
hadronization of polarized quarks are predicted
\cite{Liang:2004ph} independently of the hadronization scenarios.
Measurements of such global hyperon polarization are made possible
by the self-analyzing power of the hyperon's parity-violating
decay \cite{tdlee}, which produces an angular distribution,
$dN/d\cos\theta=1+\alpha P_H\cos\theta$, for its decay products
with respect to the polarization direction. Here, $\alpha$ is a
constant ({\it e.g.}, $\alpha=0.642$ for $\Lambda\rightarrow
p\pi^-$) and $P_H$ is the hyperon's polarization. In practice, one
needs to know  not only the orientation  but also the sign (or
direction) of the reaction plane which can only be determined by
the directed flow in each event class. If one sums over events
with opposite reaction planes, the effect of the hyperon's
polarization on the angular distribution of its decay products
will cancel. While such measurements are underway, other
consequences can and should be studied.

In this note, we discuss the spin alignment of vector mesons
due to the global quark polarization in  non-central $A+A$
collisions. Similarly as in Ref.~\cite{Liang:2004ph},
we consider two colliding nuclei with the projectile of
beam momentum $\vec{p}$ moving in the direction of the $z$ axis.
The impact parameter $\vec{b}$ is taken as along $\hat{x}$,
which is the transverse distance of the projectile from the
target nucleus. The norm of the reaction plane is given by
$\vec{n}_b\equiv \vec{p} \times \vec{b}/|\vec{p}\times\vec{b}|$
and is along $\hat{y}$. For a non-central collision, the
dense matter produced in the overlapped region of the collision
will carry a global angular momentum along the direction opposite to
the reaction plane ($-\hat{y}$).
Assuming that a partonic system is formed immediately following the
initial collision, interactions among produced partons will
lead to formation of a quark-gluon plasma (QGP) with both
transverse (in $x$-$y$ plane) and longitudinal
collective motion. The global orbital angular momentum of
the system will result in finite transverse (along $\hat{x}$)
gradient of the longitudinal flow velocity.
Given the range of interaction $\Delta x$, two colliding
partons will have relative longitudinal momentum
$\Delta p_z=\Delta x dp_z/dx$ with orbital angular momentum
$L_y \sim -\Delta x\Delta p_z$ along the direction of $\vec{n}_b$.
$L_y$ is estimated \cite{Liang:2004ph} to be of the order of one
for semi-central $Au+Au$ collisions at $\sqrt{s}=200$ GeV
and $\Delta x=1$ fm.

Such local relative orbital angular momentum $L_y$
will lead to global quark polarization via parton scattering in
QGP due to spin-orbital coupling.
The global quark polarization via elastic scattering
was calculated in an effective static potential
model and was found \cite{Liang:2004ph},
\begin{equation}
P_q = - \frac{\pi}{4}\frac{\mu\;p}{E(E+m_q)} \, ,
\end{equation}
where $E$ and $p$ are the energy and momentum of the initial quark
in the c.m. frame of the parton scattering, $\mu$ is the Debye
screening mass of the quark in medium, which specifies the average
interaction range. The polarization can be enhanced by multiple
scattering. Hence, we expect a significant global polarization of
quarks and anti-quarks before hadronization. Such global
polarization of quarks and anti-quarks will lead not only to the
global hyperon polarization but also spin alignment of vector
mesons.

The alignment of a vector meson is
described by the spin density matrix $\rho$ or its element $\rho_{m,m'}$,
where $m$ and $m'$ label the spin component along the quantization axis.
The diagonal elements $\rho_{11}$, $\rho_{00}$ and $\rho_{-1-1}$
for the unit-trace matrix are the relative intensities of the meson spin
component $m$ to take the values $1$, $0$, and $-1$ respectively,
which should be $1/3$ in the unpolarized case. Since vector mesons
usually decay strongly into two pseudo-scalar mesons,
it is difficult to measure all the elements of $\rho$.
But some of them can be determined easily by measuring
the angular distributions of the decay products.
It can be shown that, in the rest frame of $V$,
for the decay $V\to h+h'$, (where $h$ and $h'$ are two
pseudo-scalar mesons),
the angular distribution
$W(\theta,\phi)\equiv dN/d\Omega$
of the decay products is given by \cite{Schilling:1969um}
\begin{eqnarray}
W(\theta,\phi)&=&\frac{3}{4\pi}\{
\cos^2\theta \;\rho_{00}+\sin^2\theta\;(\rho_{11}+\rho_{-1-1})/2
\nonumber \\
&&-\sin2\theta (\cos\phi {\rm Re}{\rho_{10}}-\sin\phi {\rm Im}{\rho_{10}})/\sqrt{2}
\nonumber \\
&&+\sin2\theta (\cos\phi {\rm Re}{\rho_{-10}}+\sin\phi {\rm Im}{\rho_{-10}})/\sqrt{2}
\nonumber \\
&&-\sin^2\theta[\cos(2\phi){\rm Re}{\rho_{1-1}}-\sin(2\phi) {\rm Im}{\rho_{1-1}}]\}. \nonumber \\
\end{eqnarray}
Here $\theta$ is the polar angle between the direction of motion of $h$
and the quantization axis, $\phi$ is the azimuthal angle.
By integrating over $\phi$, we obtain,
\begin{equation}
W(\theta)=\frac{3}{4}
[(1-\rho_{00})+(3\rho_{00}-1)\cos^2\theta]. \label{eq-theta}
\end{equation}
Similarly, by integrating over $\theta$, we obtain,
\begin{equation}
W(\phi)=\frac{1}{2\pi}
[1-2\cos(2\phi){\rm Re}{\rho_{1-1}}+2\sin(2\phi) {\rm Im}{\rho_{1-1}}].
\end{equation}
We see that a deviation of $\rho_{00}$ from $1/3$ will lead to
a non-uniform $\theta$-distribution of the decay product. By
measuring $W(\theta)$, we can determine $\rho_{00}$.
Other elements, $\rho_{10}$ and $\rho_{1-1}$,
can be studied by further measuring $W(\theta,\phi)$.
In fact, such measurements have already been carried out in
lepton induced reactions and hadron-hadron
collisions \cite{Chliapnikov:1972ei,Azhinenko:1980cr,
Paler:1975qf,Arestov:1980mb,Cohen:1980zg,Barth:1982td,
Wittek:1987ha,Zaetz:1995vg,Aleev:2000tb,Abreu:1997wd,
Ackerstaff:1997kj,Ackerstaff:1997kd,Abbiendi:1999bz}.

Unlike the polarization of hyperons, the spin-alignment
of vector mesons, $\rho_{00}^V$, does not know the
direction of the reaction plane since it only depends
on $\cos^2\theta$ [see Eq.~(\ref{eq-theta})].
Therefore, one cannot measure the sign of the quark
polarization through spin-alignment of vector mesons.
On the other hand, one does not need to determine the
direction of the reaction plane to measure the spin alignment
which is directly related to the magnitude of the quark
polarization along the orientation of the reaction plane.

We now assume that quarks and anti-quarks
in the QGP are polarized as described 
in \cite{Liang:2004ph} and
calculate the spin alignment of $V$
by considering the following three different hadronization scenarios:
(1) recombination of the polarized quarks and anti-quarks;
(2) recombination of the polarized quarks (anti-quarks)
    with unpolarized anti-quarks (quarks);
(3) fragmentation of polarized quarks (or anti-quarks).

The picture envisaged here is the following. 
In a non-central $A+A$ collision, a QGP is formed and  
the quarks and anti-quarks in the QGP are polarized. 
Besides them, there are also quarks and anti-quarks 
created in the accompanying processes such as 
the hard scattering of the partons and the subsequent parton cascade etc. 
These quarks and anti-quarks are characterized by higher 
transverse momenta and are unpolarized. 
Hence, there are different possibilities for hadrons to be produced. 
First, they can be produced via the recombination of  
the quarks and anti-quarks in QGP, this corresponds to the 
hadronization scenario (1). 
Second, they can also be formed via the 
recombination of the quarks/anti-quarks in QGP 
with those from the accompanying processes. 
In this case, we have the recombination of polarized 
quarks (anti-quarks) with unpolarized anti-quarks (quarks), 
and this corresponds to the hadronization scenario (2).
Finally, they can also be produced via the fragmentation of a fast  
quark/anti-quark from the QGP. 
This corresponds to the scenario (3). 
Clearly, the three different hadronization 
scenarios should contribute to different kinematic regions. 
While the first scenario should play the dominant role 
in the low $p_T$ and central rapidity region, the second 
and third should play the important roles 
for the intermediate $p_T$ and forward rapidity regions respectively.

We first consider the hadronization scenario (1) of
constituent quark recombination in which both
quarks and anti-quarks are polarized.
This is likely the case for hadronization in the
central rapidity region for low $p_T$ hadrons.
We take $-\vec n_b=-\hat{y}$ as the quantization axis,
and obtain the spin density matrix for quarks $\rho^{q}$
as,
\begin{equation}
\rho^{q}= \frac{1}{2}
\left(
\begin{matrix}
&1+P_{q} &  0        \cr
&0       &   1-P_{q} \cr
\end{matrix}
\right),    \label{eq:q-pol}
\end{equation}
and similarly for anti-quarks $\rho^{\bar q}$.
Since the system is thermalized, 
there should be no intrinsic correlation between 
the quark and anti-quark in QGP. 
Also, since our purpose is to study the effect of global 
quark polarization, we will not go to the detail of the 
recombination mechanism but, 
just as people usually do\cite{Hwa:2003zp,Greco:2003xt},
assume no particular correlation between the 
quark and the anti-quark that combine into a vector meson.
Hence, we can calculate the spin density matrix of
the vector meson $V$ by making the direct product
of $\rho^{q}$ and $\rho^{\bar q}$.
After transforming it to the coupled basis, we obtain the
normalized spin density matrix $\rho^V$ for vector mesons as,
\begin{equation}
\rho^{V}= \left (
\begin{matrix}
\frac{(1+P_q)(1+P_{\bar q})}{3+P_qP_{\bar q}} & 0&0\\
0& \frac{1-P_qP_{\bar q}}{3+P_qP_{\bar q}} & 0\\
0&0& \frac{(1-P_q)(1-P_{\bar q}}{3+P_qP_{\bar q}})
\end{matrix}  \right).
\end{equation}
Hence, we obtain
\begin{equation}
\rho_{00}^{V({\rm rec})}=\frac{1-P_qP_{\bar q}}{3+P_qP_{\bar q}},
\label{eq:align}
\end{equation}
and all the non-diagonal elements are zero.
Assuming $P_u=P_d=P_{\bar u}=P_{\bar d}\equiv P_q$, and
$P_s=P_{\bar s}$, we obtain the results
for $\rho$ and $K^*$ mesons as,
\begin{equation}
\rho^{\rho({\rm rec})}_{00}=\frac{1-P_q^2}{3+P_q^2},
\end{equation}
\begin{equation}
\rho^{K^*({\rm rec})}_{00}=\frac{1-P_qP_s}{3+P_qP_s},
\end{equation}
We see that both $\rho^{\rho}_{00}$ and $\rho^{K^*}_{00}$
are smaller than $1/3$ if they are produced via
recombination of similarly polarized quarks and anti-quarks.
The non-diagonal elements are zero if there is no correlation
between the polarization of the quark and anti-quark.

The polarization of quark and anti-quark 
discussed in \cite{Liang:2004ph} is a low $p_T$ phenomenon,
since the polarizing interaction typically has a momentum scale
of $p_0=\mu L_0$, where $1/\mu$ is the interaction range and $L_0$ is
the typical relative orbital angular momentum between two-colliding
partons. When the initial $p_T$ of a quark is much larger than $p_0$,
the quark will not be polarized. But such a quark can still recombine
with a polarized low $p_T$ anti-quark to form a hadron,
according to the hadronization scenario (2).
The spin alignment for such formed vector mesons can be obtained
by inserting $P_q=0$ or $P_{\bar q}=0$ into Eq.~(\ref{eq:align}).
We have then $\rho_{00}^V=1/3$, even if one of the constituent
quarks is polarized before recombination.

Finally, we consider the hadronization scenario (3),
{\it i.e.}, fragmentation of a polarized quark $q^\uparrow\to V+X$.
This likely happens for quarks with large rapidities in the QGP  
and may play an important for hadrons in the forward rapidity region. 
The situation in this case 
is very much different from that in scenario (1) or (2). 
Here, the anti-quark that combines with the initial polarized quark 
is created in the fragmentation process and may carry the 
information of the initial quark that induces this creation. 
This implies that the polarization of this anti-quark 
can be correlated to that of the initial quark. 
Since this is a non-perturbative process that cannot be calculated 
from pQCD, we do not know {\it a priori} whether such a correlation 
indeed exists. 
Fortunately, the situation here is very similar to 
$e^+e^-\to Z^0\to q\bar q\to V+X$,
where the initial $q$ and $\bar q$
are longitudinally polarized so that 
we have the fragmentation process $\vec q\to V+X$. 
Therefore, we can compare it with the latter to 
extract some useful information. 
 
The $00$-element of the spin density matrix
for the vector mesons in $e^+e^-\to Z^0\to q\bar q\to V+X$
have been measured at
LEP\cite{Abreu:1997wd,Ackerstaff:1997kj,Ackerstaff:1997kd,Abbiendi:1999bz}.
The results show clearly that $\rho^V_{00}$
is significantly larger than $1/3$
in the helicity frame of the vector meson
({\it i.e.} the quantization axis is taken as the polarization
direction of the fragmenting quark)
at large fractional momenta.
A simple calculation \cite{Xu:2001hz} for $\rho_{00}^V$ in
$e^+e^-\to V+X$ has been carried out
by building the direct product of the spin density matrix
of the polarized leading quark ($\rho_q$) and that of the
anti-quark created during the fragmentation
process ($\rho_{\bar q}^{\rm frag}$). In the helicity frame, $\rho_q$
takes exactly the form as shown by Eq.~(\ref{eq:q-pol}).
The most general form was taken for $\rho_{\bar q}^{\rm frag}$.
The calculation is exactly the same as that for quark
recombination. It also leads to a result of $\rho_{00}^V$
for the first rank $V$'s similar to that shown by Eq.~(\ref{eq:align}).
The only difference is that we should replace
$P_{\bar q}$ in Eq.~(\ref{eq:align}) by $P_{\bar q}^{frag}$,
which is the polarization of the anti-quark created in the
fragmentation process. This result has been compared with the available
data\cite{Abreu:1997wd,Ackerstaff:1997kj,Ackerstaff:1997kd,Abbiendi:1999bz}.
It has been found out that, the available data can only be fitted if
the anti-quark is taken as effectively polarized in
the opposite direction as the leading quark, and
the polarization is $P_{\bar q}^{\rm frag}=-\beta P_q$,
where $\beta\approx 0.5$ was obtained\cite{Xu:2001hz} by fitting the
data\cite{Abreu:1997wd,Ackerstaff:1997kj,Ackerstaff:1997kd,Abbiendi:1999bz}.
Hence, for the first rank $V$'s,
\begin{equation}
\rho^{V(frag)}_{00}=\frac{1+\beta P_q^2}{3-\beta P_q^2}.
\label{eq:frg}
\end{equation}
For $V$'s other than the first rank hadrons, $\rho^V=1/3$.
These results can be considered as
a parametrization of the LEP
data\cite{Abreu:1997wd,Ackerstaff:1997kj,Ackerstaff:1997kd,Abbiendi:1999bz}.

If the same model can be applied to the fragmentation
of quarks (anti-quarks) polarized along the opposite direction of the
reaction plane in heavy-ion collisions, then the anti-quarks (quarks)
that are produced in the fragmentation and will combine with the
leading quarks (anti-quarks) to form vector mesons
is effectively polarized in the opposite direction as the
initial quarks (anti-quarks) with the polarization
$P_{\bar q}^{\rm frag}=-\beta P_q$.
One can then obtain a result for $\rho_{00}^{V}$ in
the same form as that shown by Eq.~(\ref{eq:frg}).
The difference is that now the quantization axis is along
the opposite direction of the reaction plane, which
is transverse to the direction of longitudinal motion.
Taking the fragmentation of different flavors of
quarks and anti-quarks into account,
we obtain, for the first rank $V$'s,
\begin{equation}
\rho^{\rho\ ({\rm frag})}_{00}=\frac{1+\beta P_q^2}{3-\beta P_q^2},
\end{equation}
\begin{equation}
\rho^{K^*({\rm frag})}_{00}=\frac{f_s}{n_s+f_s}\frac{1
+\beta P_q^2}{3-\beta P_q^2}+
\frac{n_s}{n_s+f_s}\frac{1+\beta P_s^2}{3-\beta P_s^2},
\end{equation}
where $n_s$ and $f_s$ are the strange quark abundances relative to
up or down quarks in QGP and quark fragmentation, respectively.
Therefore, in this case of quark fragmentation, $\rho_{00}$ is
always larger than $1/3$.

One can measure directly the angular distribution of vector
mesons' decay products with respect to the reaction plane and
therefore determine the spin-alignment of vector mesons in
non-central heavy-ion collisions. Such measurements will
elucidate the hadronization mechanisms in the different
kinematic regions if $\rho_{00}$ differs noticeably from 1/3.
Before such data become available, one can, however, find
constraints on $\rho_{00}$ from the measured azimuthal anisotropy
of produced hadrons, since finite fraction of final hadrons come
from vector meson decays and they have a particular angular
distribution with respect to the reaction plane according to
Eq.~(\ref{eq-theta}) if $\rho_{00}\not=1/3$. Such an angular
distribution will produce an azimuthal asymmetry with respect
to the reaction plane. If one characterizes the asymmetry by
the second coefficient $v_2$ of the Fourier transformation of
the angular distribution similarly as the elliptic flow
study \cite{eflow}, $v_2>0$ for $\rho_{00}<1/3$ and $v_2<0$
if $\rho_{00}>1/3$.

Note that the angular distribution in Eq.~(\ref{eq-theta})
is in the rest frame of the decaying vector mesons. To
calculate the azimuthal anisotropy of the decay products
in the center of mass (c.m.) frame of $A+A$ collisions, one
has to perform Lorentz transformation on the momentum
distribution of the decay products from the rest frame of the
vector mesons. Since quarks' polarizations should disappear
at large $p_T$ as we have argued earlier, the vector mesons'
spin alignment should approach to $\rho_{00}=1/3$ at
large $p_T$. Therefore, we can assume the following ansatz
for the $p_T$ dependence of $\rho_{00}$,
\begin{equation}
\rho_{00}(p_T)=\rho_{00}^0
+(\frac{1}{3}-\rho_{00}^0)\frac{2}{\pi}\tan^{-1}(\frac{p_T}{a_0}),
\label{eq:r00}
\end{equation}
where $\rho_{00}^0$ is the value of the spin alignment at $p_T=0$ and
$a_0$ sets the $p_T$ scale at which quark's polarization vanishes.
We will use $a_0=0.5$ GeV/$c$ to illustrate the effect of vector
mesons' spin alignment on the $v_2$ of final hadrons.

\begin{figure}[htb!]
\resizebox{3.0in}{2.0in}{\includegraphics{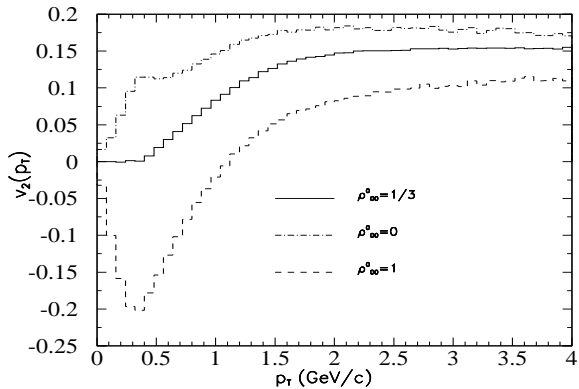}}
\caption{Azimuthal anisotropy $v_2$ of pions from the decay of
$\rho$ vector mesons that have spin
alignment according to Eq.~(\protect\ref{eq:r00}) with
$\rho_{00}^0=1/3$ (solid line), 0 (dot-dashed line)
and 1 (dashed line).\label{fig1}}
\end{figure}

Effects of vector resonances' decay on final pions' azimuthal
asymmetry has been studied recently \cite{nxu,ko} without
spin alignment. We will follow the same procedure and assume
that vector mesons at low $p_T$ have an exponential distribution
in $m_T=\sqrt{p_T^2+m^2}$ with an effective temperature $T=200$ MeV.
The azimuthal anisotropy of $\rho$ mesons is assumed to follow the
scaling behavior of a recombination model \cite{nxu,ko}
\begin{equation}
v_2^{\rho}=\frac{0.22}{1.0+e^{-(p_T/2.0-0.35)/0.2}}-0.06.
\end{equation}
Shown in Fig.~\ref{fig1} are the azimuthal anisotropies of
pions from $\rho$ meson decays with three limiting cases
of $\rho$-mesons' spin alignment. For $\rho_{00}^0=1/3$, $\rho$-mesons
are not aligned, pion distribution in the rest frame of
the $\rho$-meson is isotropic. Therefore, $v_2$ of pions
(solid line) from the decay follows  closely that of the $\rho$
mesons. For one extreme case, $\rho_{00}^0=1$,
the angular distribution of pions
prefers the out-plane direction and therefore $v_2$ is negative
at low $p_T$ shown as dashed line.
As the transverse momentum of the $\rho$-meson
increases, the opening angle of pions from the decay in the c.m.
frame of $A+A$ collisions becomes smaller. Eventually, $v_2$
of pions approaches that of $\rho$-meson (solid line) at high $p_T$.
But it is always smaller than $v_2$ of the $\rho$-mesons.
For another extreme case, $\rho_{00}^0=0$, the azimuthal
anisotropy of pions (dot-dashed line) is positive and larger
than that of the $\rho$-mesons for small $p_T$ while it
approaches to the $\rho$-mesons' $v_2$ at large $p_T$ from above.

\begin{figure}[htb!]
\resizebox{3.0in}{2.0in}{\includegraphics{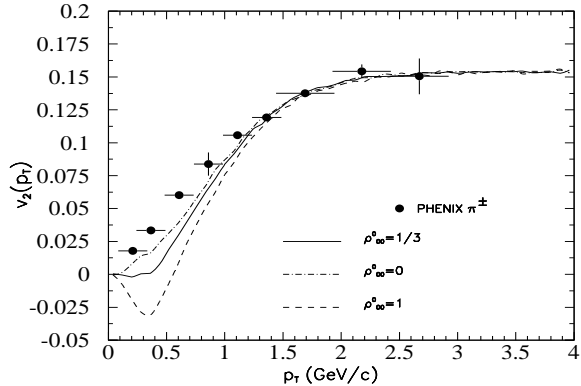}}
\caption{Azimuthal anisotropy $v_2$ of final produced
pions that include both directly produced and these from
the decay of $\rho$ vector mesons that have spin
alignment according to Eq.~(\protect\ref{eq:r00}) with
$\rho_{00}^0=1/3$ (solid line), 0 (dot-dashed line)
and 1 (dashed line). The multiplicity
ratio $\rho/\pi=0.183$ \protect\cite{star-v} is assumed.
The data are for pions from PHENIX
experiment \protect\cite{phenix}.\label{fig2}}
\end{figure}

Since there are also directly produced pions, one needs to
also include them in the
estimate of the effect of $\rho$-mesons' spin alignment on
the azimuthal anisotropy of the final pions. The $\rho/\pi$
ratio is measured to be 0.183 in peripheral $Au+Au$ collisions
at $\sqrt{s}=200$ GeV about the same value as in $p+p$ at the
same energy \cite{star-v}. We simply assume the same value for
the total number of $\rho$ mesons. We also assume that $v_2$ for the
directly produced pions is the same as $\rho$-mesons. Shown
in Fig.~\ref{fig2} are $v_2(p_T)$ of the final produced
pions with the above three limiting cases of $\rho$-mesons'
spin alignment as compared to the experimental data \cite{phenix}
which are always above the assumed $v_2$ of the $\rho$-mesons
(solid line) in the scaling model \cite{nxu,ko}.
The effect of vector mesons' spin alignment is not big due
to the small value of $\rho/\pi$ ratio. In addition to pions
from $\rho$-meson decays, one also finds \cite{nxu,ko} that
pions from decays of other resonances such as $\Delta$
and $\omega$ also enhances pions' $v_2$ at low $p_T$.
For a complete analysis, one should include decays of these
resonances that might also have similar spin alignment.
However, it is clear that the data favor the scenario of
$\rho_{00}^0<1/3$ from parton recombination, even though
the measured $v_2$ of pions does not provide stringent
constraints on the value of $\rho$-mesons' spin alignment,
which can be directly measured by the angular distribution
of the decay products in the rest frame of vector mesons.


In summary, we calculated the spin density matrix for
vector mesons produced in non-central $A+A$ collisions
by taking into account the global polarization of
quarks and anti-quarks in the overlapping region of
the two colliding nuclei. We showed that the results for $\rho_{00}^V$
depend very much on the hadronization mechanisms.
Since mesons from the decay of aligned vector resonances
have anisotropic angular distribution with respect to the
reaction plane, they will contribute to the azimuthal
anisotropy $v_2$ of the final produced meson spectra. Even
though experimental data on pions' $v_2$ cannot provide stringent
constraints on the value of $\rho$-mesons' spin alignment,
they favor $\rho_{00}^0<1/3$ for the scenario of
parton recombination in the central rapidity region.
For the parton fragmentation scenario, $\rho_{00}^0>1/3$
and $v_2$ of the decay products is negative. In this case
the spin-alignment tends to decrease the $v_2$ of the total
produced mesons. Such scenario is more likely in the large
rapidity region and can potentially explain why the measured
$v_2$ drops sharply with rapidity.

It is also interesting to note that similar quark polarization
effect could also happen in $p+p$ collisions as pointed
out in Ref.~\cite{Voloshin:2004ha} and could explain the
measured hyperon polarization. Therefore, it should
also be interesting to correlate hyperon's polarization
to the reaction plane in $p+p$ collisions if one can
similarly determine the reaction plane as in $A+A$ collisions.
However, one  must also take into account the
spin of partons inside initial beam proton as one cannot
neglect them in $p+p$ collisions.

This work was supported by DOE
the Director, Office of Energy
Research, Office of High Energy and Nuclear Physics, Divisions of
Nuclear Physics, of the U.S. Department of Energy
under No. DE-AC03-76SF00098,
National Natural Science Foundation of China
NSFC No. 10175037 and No. 10440420018.



\begin{thebibliography}{99}
\bibitem{Liang:2004ph}
Z.~T.~Liang and X.~N.~Wang,
Phys. Rev. Lett. {\bf 94}, 102301 (2005),
arXiv:nucl-th/0410079.

\bibitem{tdlee}
T.~D.~Lee and C.~N.~Yang,
Phys.\ Rev.\  {\bf 108}, 1645 (1957).
T.~D.~Lee, J.~Steinberger, G.~Feinberg, P.~K.~Kabir and C.~N.~Yang,
Phys.\ Rev.\  {\bf 106}, 1367 (1957).



\bibitem{Schilling:1969um}
K.~Schilling, P.~Seyboth and G.~E.~Wolf,
Nucl.\ Phys.\ B {\bf 15}, 397 (1970)
[Erratum-ibid.\ B {\bf 18}, 332 (1970)].

\bibitem{Chliapnikov:1972ei}
P.~Chliapnikov, O.~Czyzewski, Y.~Goldschmidt-Clermont, M.~Jacob and P.~Herquet,
%
Nucl.\ Phys.\ B {\bf 37}, 336 (1972).

\bibitem{Azhinenko:1980cr}
I.~V.~Azhinenko {\it et al.}  [French-Soviet Collaboration],
%
Z.\ Phys.\ C {\bf 5}, 177 (1980).

\bibitem{Paler:1975qf}
K.~Paler {\it et al.},
%
Nucl.\ Phys.\ B {\bf 96}, 1 (1975).

\bibitem{Arestov:1980mb}
Y.~Arestov {\it et al.}  [French-Soviet Collaboration],
%
Z.\ Phys.\ C {\bf 6}, 101 (1980).

\bibitem{Cohen:1980zg}
I.~Cohen {\it et al.},
Phys.\ Rev.\ D {\bf 25}, 634 (1982).

\bibitem{Barth:1982td}
M.~Barth {\it et al.}  [Brussels-Genoa-Mons-Nijmegen-Serpukhov-CERN
                  Collaboration],
Nucl.\ Phys.\ B {\bf 223}, 296 (1983)
[Erratum-ibid.\ B {\bf 232}, 547 (1984)].

\bibitem{Wittek:1987ha}
W.~Wittek {\it et al.}  [BEBC WA59 Collaboration],
%
Phys.\ Lett.\ B {\bf 187}, 179 (1987).

\bibitem{Zaetz:1995vg}
V.~G.~Zaetz {\it et al.}  [Big Bubble Chamber Neutrino Collaboration],
%
Z.\ Phys.\ C {\bf 66}, 583 (1995).

\bibitem{Aleev:2000tb}
A.~N.~Aleev {\it et al.}  [EXCHARM Collaboration],
%
Phys.\ Lett.\ B {\bf 485}, 334 (2000)
[arXiv:hep-ex/0002054].

\bibitem{Abreu:1997wd}
P.~Abreu {\it et al.}  [DELPHI Collaboration],
%
Phys.\ Lett.\ B {\bf 406}, 271 (1997).

\bibitem{Hwa:2003zp}
  R.~C.~Hwa and C.~B.~Yang,
  Phys.\ Rev.\ Lett.\  {\bf 90}, 212301 (2003)
  [arXiv:nucl-th/0301004].

\bibitem{Greco:2003xt}
  V.~Greco, C.~M.~Ko and P.~Levai,
  Phys.\ Rev.\ Lett.\  {\bf 90}, 202302 (2003)
  [arXiv:nucl-th/0301093].

\bibitem{Ackerstaff:1997kj}
K.~Ackerstaff {\it et al.}  [OPAL Collaboration],
Phys.\ Lett.\ B {\bf 412}, 210 (1997)
[arXiv:hep-ex/9708022].

\bibitem{Ackerstaff:1997kd}
K.~Ackerstaff {\it et al.}  [OPAL Collaboration],
Z.\ Phys.\ C {\bf 74}, 437 (1997).

\bibitem{Abbiendi:1999bz}
G.~Abbiendi {\it et al.}  [OPAL Collaboration],
%
Eur.\ Phys.\ J.\ C {\bf 16}, 61 (2000)
[arXiv:hep-ex/9906043].



\bibitem{Xu:2001hz}
Q.~H.~Xu, C.~X.~Liu and Z.~T.~Liang,
Phys.\ Rev.\ D {\bf 63}, 111301 (2001)
[arXiv:hep-ph/0103267].

\bibitem{eflow}
A.~M.~Poskanzer and S.~A.~Voloshin,
Phys.\ Rev.\ C {\bf 58}, 1671 (1998)
[arXiv:nucl-ex/9805001].

\bibitem{nxu}
X.~Dong, S.~Esumi, P.~Sorensen, N.~Xu and Z.~Xu,
Phys.\ Lett.\ B {\bf 597}, 328 (2004)
[arXiv:nucl-th/0403030].

\bibitem{ko}
V.~Greco and C.~M.~Ko,
Phys.\ Rev.\ C {\bf 70}, 024901 (2004)
[arXiv:nucl-th/0402020].

\bibitem{phenix}
S.~S.~Adler {\it et al.}  [PHENIX Collaboration],
Phys.\ Rev.\ Lett.\  {\bf 91}, 182301 (2003)
[arXiv:nucl-ex/0305013].


\bibitem{star-v}
J.~Adams {\it et al.}  [STAR Collaboration],
Phys.\ Rev.\ Lett.\  {\bf 92}, 092301 (2004)
[arXiv:nucl-ex/0307023].

\bibitem{Voloshin:2004ha}
S.~A.~Voloshin,
arXiv:nucl-th/0410089.


\end{thebibliography}
\end{document}